\documentclass[journal]{IEEEtran}

\usepackage[latin1]{inputenc}
\usepackage{times,amsmath}
\usepackage{amssymb}
\usepackage{steinmetz}
\usepackage{pstool}
\usepackage{multirow}
\usepackage{enumerate}
\usepackage{graphicx}
\usepackage{multirow}
\usepackage{subcaption}
\usepackage{mwe}
\usepackage{bigints}
\usepackage{MnSymbol}
\usepackage{stfloats} 
\usepackage[table]{xcolor}
\usepackage[square, comma, sort&compress, numbers]{natbib}
\usepackage{nohyperref}
\usepackage{algorithm,algorithmic}
\usepackage{epstopdf}
\usepackage{mathtools, cuted}
\graphicspath{ {Figures/} }

\usepackage{colortbl}
\usepackage{multirow}
\usepackage{color}
\usepackage{algorithmic}
\usepackage{graphicx}
\usepackage{textcomp}
\usepackage{wrapfig}
\usepackage{subcaption}
\usepackage{epstopdf}
\usepackage{epsfig}
\usepackage{amsmath,amssymb,amsfonts}

\usepackage{comment}

\definecolor{pastelblue}{rgb}{0.85, 0.92, 0.97}

\begin{document}

\title{ Evaluation of a Low-Cost Single-Lead ECG Module for Vascular Ageing Prediction and Studying Smoking-induced Changes in ECG }

\author{
\IEEEauthorblockN{
Syed Anas Ali\IEEEauthorrefmark{1},
Muhammad Saqib Niaz\IEEEauthorrefmark{1}, Mubashir Rehman 
\IEEEauthorrefmark{1}, Ahsan Mehmood\IEEEauthorrefmark{1}, M. Mahboob Ur Rahman\IEEEauthorrefmark{1}, Kashif Riaz\IEEEauthorrefmark{1},\IEEEauthorrefmark{2}, Qammer H. Abbasi\IEEEauthorrefmark{3},\IEEEauthorrefmark{4} }
\thanks{This work was supported in part by the UK Engineering and Physical Sciences Research Council (EPSRC) grants: EP/X040518/1 and EP/T021020/1.}
\\
\IEEEauthorblockA{\IEEEauthorrefmark{1} Electrical engineering department, Information Technology University, Lahore 54000, Pakistan\\ 
\IEEEauthorrefmark{2}College of Science and Engineering, Hamad Bin Khalifa University, Qatar Foundation, Doha, Qatar\\
\IEEEauthorrefmark{3}James Watt School of Engineering, University of Glasgow, Glasgow, G12 8QQ, UK\\
\IEEEauthorrefmark{4}Artificial Intelligence Research Centre, Ajman University, Ajman, UAE\\
\IEEEauthorrefmark{1}\{msee19009,  mahboob.rahman\}@itu.edu.pk, \IEEEauthorrefmark{3}Qammer.Abbasi@glasgow.ac.uk }
}

\maketitle

\begin{abstract} 

Vascular age is traditionally measured using invasive methods or through 12-lead electrocardiogram (ECG). This paper utilizes a low-cost single-lead (lead-I) ECG module to predict the vascular age of an apparently healthy young person. In addition, we also study the impact of smoking on ECG traces of the light-but-habitual smokers. We begin by collecting (lead-I) ECG data from 42 apparently healthy subjects (smokers and non-smokers) aged 18 to 30 years, using our custom-built low-cost single-lead ECG module, and anthropometric data, e.g., body mass index, smoking status, blood pressure, etc. Under our proposed method, we first pre-process our dataset by denoising the ECG traces, followed by baseline drift removal, followed by z-score normalization. Next, we create another dataset by dividing the ECG traces into overlapping segments of five-second duration. We then feed both segmented and unsegmented datasets to a number of machine learning models, a 1D convolutional neural network, and ResNet18 model, for vascular ageing prediction. We also do transfer learning whereby we pre-train our models on a public PPG dataset, and later, fine-tune and evaluate them on our unsegmented ECG dataset. The random forest model outperforms all other models and previous works by achieving a mean squared error (MSE) of 0.07 and coefficient of determination $R^2$ of 0.99, MSE of 3.56 and $R^2$ of 0.26, MSE of 0.99 and $R^2$ of 0.87, for segmented ECG dataset, for unsegmented ECG dataset, and for transfer learning scenario, respectively. Finally, we utilize the explainable AI framework to identify those ECG features that get affected due to smoking. This work is aligned with the sustainable development goals 3 and 10 of the United Nations which aim to provide low-cost but quality healthcare solutions to the unprivileged. This work also finds its applications in the broad domain of forensic science.

\end{abstract}


\section{Introduction}

\label{sec:intro}

\begin{table}[h!]
\centering
\renewcommand{\arraystretch}{1.3}
\begin{tabular}{|l|l|}
\hline
\textbf{Acronym} & \textbf{Definition} \\ \hline
ECG & Electrocardiogram \\ \hline
CVD & Cardiovascular Disease \\ \hline
UN & United Nations \\ \hline
SDG & Sustainable Development Goal \\ \hline
WHO & World Health Organization \\ \hline
AI & Artificial Intelligence \\ \hline
ML & Machine Learning \\ \hline
DL & Deep Learning \\ \hline
TL & Transfer Learning \\ \hline
xAI & Explainable AI \\ \hline
PPG & Photoplethysmography \\ \hline
LR & Linear Regression \\ \hline
RR & Ridge Regression \\ \hline
DT & Decision Tree \\ \hline
RF & Random Forest \\ \hline
CNN & Convolutional Neural Network \\ \hline
MSE & Mean Squared Error \\ \hline
R2 & Coefficient of Determination \\ \hline
\end{tabular}
\caption{List of Acronyms}
\label{tab:acronyms}
\end{table}

Vascular age (also known as heart age or ECG-age) refers to the natural deterioration in structure and function of blood vessels as a person gets older \cite{A11}. In other words, as blood vessels age, they become stiffer, less elastic, and more prone to plaque buildup. Vascular ageing results from factors such as ageing, high blood pressure, high cholesterol, and smoking. Vascular ageing is a major risk factor for a number of cardiovascular diseases (CVD) and other diseases, including the following: hypertension (high blood pressure), atherosclerosis (narrowing and hardening of arteries), stroke, heart attacks, heart failure, endothelial dysfunction (impaired function of the inner lining of blood vessels), organ damage (due to reduced blood supply to organs, e.g., kidneys, eyes, etc.), and cognitive decline (due to reduced blood flow to the brain) \cite{A12}, \cite{A13}.

Among various risk factors, smoking is one significant factor that impacts the vascular age, and thus, impacts the cardiovascular health. Chronic smoking could lead to various alterations in the electrocardiogram (ECG), e.g., increased heart rate (tachycardia), increased risk of developing cardiac arrhythmias (such as premature atrial contractions and premature ventricular contractions), changes in the QT interval and ST-segment on the ECG (which may indicate myocardial ischemia, i.e., insufficient blood flow to the heart muscle), increased risk of coronary artery disease (shown by ST-segment depression and T-wave inversion), vasoconstriction (narrowing of blood vessels), and reduced oxygen levels \cite{smokingecg}. 

As discussed above, vascular age and smoking both result in alterations in the ECG \cite{smokingecg}, \cite{A14}. This motivates us to study the feasibility of using ECG as a clinical tool to assess the impact of vascular age and smoking on cardiovascular health. This objective is inline with the broad research efforts that aim to use ECG as a diagnostic tool with the ultimate objective of bringing down the annual CVD-based deaths by early detection of CVDs \cite{A5}. A reduction in proliferation of CVDs will imply a reduction in health burden of countries in lieu of health insurance costs. However, note that manual assessment of the ECG to look for possible irregularities in the heart rhythm due to vascular ageing and smoking is time-consuming and error-prone. Hence, artificial intelligence (AI)-based methods which could automatically and reliably predict the vascular age \cite{A16} of a person (smoker or non-smoker) from the (single-lead) ECG data are the need of the hour. This is precisely the scope of this work. 

{\it Existing methods for vascular ageing prediction and the dilemma:}
Vascular age prediction has traditionally been done using invasive methods which use catheters and pressure sensors placed directly inside arteries to obtain arterial stiffness and central aortic blood pressure measurements. Recently, there has been growing interest in designing non-invasive methods, which utilize either photoplethysmography (PPG) or ECG signals for vascular ageing prediction, \cite{noninvasivevascagereview}, \cite{A16}. Among them, pulse wave velocity (PWV) method and its variants are considered the gold standard for measuring arterial stiffness. PWV method measures the speed at which the pressure wave travels along the arterial tree between two points. One important variant is carotid-femoral PWV (cfPWV) method which reflects the stiffness of the central arteries. Brachial-Ankle PWV (baPWV) method is another important variant that measures the time taken for the pressure wave to travel between the brachial and ankle arteries. Pulse wave analysis (PWA) method is another popular method that analyzes the shape of the arterial pulse wave to derive indices related to arterial stiffness and vascular function. Other than that, there are specific methods that examine the endothelial function, e.g., ultrasound-based flow-mediated dilation method, peripheral arterial tonometry to obtain reactive Hyperaemia index, etc. However, each of these methods has its drawbacks, e.g., invasive, high cost, need for trained professionals and/or specialized equipment, inconsistent results, etc.

{\bf Contributions.} 
The main contributions of this paper are three-fold:
\begin{itemize}
    \item {\it Data acquisition:} We design a low-cost single-lead ECG sensor module (with a price tag of 10 USD), and utilize it to collect (lead-I) ECG data from 42 apparently healthy subjects (male, female, smoker, non-smoker) aged 18 to 30 years. We also segment our dataset in order to get another large dataset that consists of 6,131 data points.  
    \item {\it Vascular age estimation:} We feed the two custom datasets (segmented and unsegmented) to various machine, deep and transfer learning methods, which predict the vascular age of a person using a low-cost single-lead (lead-I) ECG sensor module only. We also do transfer learning whereby we pre-train our models on a public PPG dataset, and later, fine-tune and evaluate them on our unsegmented ECG dataset. 
    \item {\it Study of smoking-induced changes in single-lead ECG:} We also do explainable AI-based feature analysis on the two custom datasets (segmented and unsegmented) in order to identify those features that get affected due to smoking.
\end{itemize}

We note that our work advances the state-of-the-art that mostly utilizes either the 12-lead ECG \cite{A14} or invasive methods (e.g., central aortic blood pressure monitoring) for vascular age estimation. 
Further, this work is also aligned with the sustainable development goals (3 and 10) of the United Nations which aim to provide low-cost but quality healthcare sensing solutions to the unprivileged populations (e.g., people in developing countries and people in remote areas). In addition, this work also finds its applications in the broad domain of forensic science.

{\bf Outline.} Section II provides a summary of the selected related work. Section III describes the proposed methodology as well as the hardware architecture of our custom low-cost single-lead ECG module. Section IV discusses the data collection process for our custom ECG dataset, the data pre-processing steps, and feature extraction process. Section V enumerates the ML, DL and TL models we have implemented. Section VI describes selected results for vascular ageing prediction and smoking-induced changes in ECG. Finally, Section VII concludes the paper.

\section{Related work}

\subsection{Vascular ageing estimation methods}

The works on vascular age estimation could be categorized into the following broad categories: statistical methods, machine learning methods, deep learning methods, and classical score based methods. In terms of hardware equipment, the non-invasive methods reported in the literature either utilize the 12-lead ECG or PPG sensors. For example, Ladejobi et. al. \cite{A14} provide a 12-lead ECG method for vascular age estimation. Similarly, Charlton et. al. \cite{ppgvascagenet} provide an in-depth and comprehensive review of the works that utilize PPG sensors to compute various parameters for the sake of vascular age estimation. Next, we consider selected works, group together the similar works under an umbrella term and discuss them below, one by one.  

{\it Statistical methods:}
Ball et. al. \cite{bbb} develop a Bayesian statistical model that predicts an individual's heart age based on his/her resting 5 min 12-lead ECG. They evaluated their model on 776 healthy individuals, 221 with cardiovascular risk factors, 441 with cardiac disease, and a small group of highly endurance-trained athletes. Their model predicts a higher heart age for the subjects with risk factors, with heart diseases, and athletes. Motivated by the situations where cfPWV cannot be measured, Heffernan et. al. \cite{7} compute the estimated pulse wave velocity (ePWV) from the age and brachial blood pressure. They measure cfPWV in two-hundred and fifty-two adults, and compare it with each participant's ePWV, in order to correlate the ePWV with cfPWV (which is an established measure of vascular aging).

{\it Machine learning based methods:}
Recently, machine learning methods for vascular ageing prediction have gained widespread attention (see the survey article \cite{A16} and references therein for a quick review of the development in the field). Due to space constraints, we discuss selected works below. Kakadiaris et. al. \cite{kakadiaris2018machine} utilizes a 13-years MESA dataset of 6459 atherosclerotic CVD-free participants and demonstrates that their proposed support vector machine (SVM)-based CVD risk calculator outperforms the classical PCE-based CVD risk calculator. Dall Olio et. al. \cite{liu2018support} utilizes a public dataset of video PPG signals acquired from smartphones, does the feature extraction, and trains a number of machine and deep learning methods in order to predict healthy vascular ageing. Suri et. al. \cite{suri2022understanding} take a critical approach and  study the bias problem in ML-based vascular age prediction methods. They argue that overfitting and lack of large datasets (with diverse ethnic populations) might be the reasons behind the fact that ML methods yield inaccurate results during the clinical trials. Similarly, Lindow et. al. \cite{aziz2021heartbeat} argues that it is possible to accurately predict patients' Bayesian 5-min ECG Heart ages from their standard, resting 10-s 12-lead ECGs using standard regression models only, without relying upon the AI techniques. They utilize a public dataset with a total of 2,771 subjects (1682 healthy volunteers, 305 with cardiovascular risk factors, 784 with cardiovascular disease). 

{\it Deep learning based methods:}
In \cite{lima2021deep}, Lima et. al. train a residual neural network on the 12-lead ECG data from the CODE study cohort (1,558,415 patients), and report that the patients with a heart age gap of more than 8 years have a higher mortality rate. Hirota et. al. \cite{urtnasan2021ai} train a convolutional neural network (CNN) on 12-lead ECG data of 17,042 subjects, and reports that the age predicted by their model matches well with the chronological age of young population, but not for the elderly (with age $\ge60$ years). Chang et. al. \cite{chang2022electrocardiogram} train the ECG12Net model (basically, a CNN with 82 convolutional layers) on 71,741 subjects (20 to 80 years old), and reports that the high ECG-age is not only correlated with mortality but with several CVD-related outcomes as well. In \cite{parkvascular}, Park et. al. collect PPG data from 757 participants, utilize a Gaussian mixture model to decompose the PPG signal into incident and reflected waves for feature extraction, and eventually train a shallow artificial neural network for vascular ageing prediction. Libiseller-Egger et. al. \cite{ince2017learned} train a CNN on 10-second 12-lead ECG data and CVD-related metadata from 499,727 subjects, with the aim to learn the genetic basis of cardiovascular ageing, in order to support future personalised/precision medicine (when genomic information is readily available). In \cite{dzz}, Attia et. al. train their CNNs using 10-second samples of 12-lead ECG signals from 499,727 patients in order to predict their biological age and gender. Benavente et. al. \cite{izz} train a CNN on over 700,000 ECGs from the Mayo Clinic, US, in order to later predict the ECG-age of 4,542 participants in two cities in Russia. Authors report that $\delta$-age (i.e., ECG-derived artificial intelligence-estimated age minus chronological age) biomarker is strongly correlated with the CVD risk factors, e.g., blood pressure, BMI, total cholesterol and smoking. In \cite{hzz}, Toya et. al. train a CNN on the 12-lead ECG data of 774,783 unique subjects, and conclude that the patients with peripheral microvascular endothelial dysfunction (an index of vascular aging) and accelerated physiologic aging are at a greater risk for CVDs.

{\it Classical CVD risk calculator methods:}
In addition to vascular ageing prediction methods, other methods with related objectives also exist. For example, the Framingham risk score (FRS) \cite{wilson1998prediction}, pooled cohort equations (PCE) \cite{gibbons1997acc}, and systematic vascular risk evaluation (SCORE) are few widely used tools that take into account various risk factors such as gender, systolic blood pressure, smoking habits, and HDL cholesterol in order to estimate an individual's 10-year risk of developing a cardiovascular event, such as coronary heart disease or stroke. 
However, it has been reported that the FRS method overestimates the risk in young age groups, while the PCE method overestimates the risk in older age groups \cite{ko2020calibration}. 

{\it Low-cost single-lead ECG module design:}
Since this work utilizes the AD8232 chip (a low-cost single-lead ECG sensor with a price tag of 5 USD), it is imperative to summarize the works that design low-cost single-lead ECG module with the unanimous aim to provide affordable sensing solutions for remote health monitoring. There are works that consider a AD8232 chip-based low-cost single-lead ECG module for atrial fibrillation detection (a common cardiac arrhythmia associated with stroke risk) \cite{ukil2022afsense},\cite{ahsanuzzaman2020low}. Gromer et. al. \cite{gromer2019ecg} present a cost-effective method for detecting driver drowsiness, utilizing heart rate variability data from the ECG sensor. Further, works do exist that focus on the design of an internet-of-things (IoT)-based low-cost ECG system which reports various biomarkers (e.g., pulse rate, SpO2, body temperature, etc) of remote Covid-19 patients to the doctor, and could detect various cardiac abnormalities such as premature ventricular contraction, premature atrial complex, etc \cite{singh2017iot,rahman2022real,yin2020influence}. Porumb et. al. \cite{porumb2020precision} aim to estimate blood glucose level in a subject using a non-invasive wearable ECG device, using features such as T-wave, ST-segment, etc. 

{Last but not the least, there are works that aim to estimate the biological age of a person using ECG or PPG data. The interested reader is referred to \cite{saranVApaper} and the references therein.}

\subsection{Effect of smoking on ECG}

According to epidemiological studies, Tobacco cigarette smoking is one of the (preventable) major risk factor for cardiovascular diseases, e.g., coronary artery disease, arrhythmia, atherosclerosis, ischemic heart diseases, and sudden cardiac death.  
This is because Nicotine, an ingredient of cigarette, alters the physiology of the heart by inducing conduction block, and by producing adverse effects on ventricular repolarization. The Nicotine also acts as a potassium channel blocker, and leads to abnormal ventricular repolarization, atrial depolarization, and atrioventricular depolarization, which in turn implies cardiac arrhythmia and more. Below, we describe selected works that have studied the impact of smoking on cardiac health through ECG waveform.

In \cite{s1}, Yadav et. al. collect ECG data from male smoker and non-smoker subjects in the age range of 18-70 years, and observe a decrease in PR interval, decrease in QRS complex duration, change in R-wave amplitude and S-wave amplitude, change in ST segment duration, decrease in TP interval, and change in T-wave duration, for heavy smokers. 
Devi. et. al. \cite{s2} collect ECG data from 88 healthy male subjects (44 smokers, according to the ICD-10 criterion) in the age group of 18-30 years, in a hospital setting. They observe a decrease in QTc interval, increase in the QRS complex duration, increase in the heart rate, and decrease in the RR interval, the QT interval and the ST segment, for the smokers. 
Kumar. et. al. \cite{s5}
collect 12-lead ECG data from 148 male patients, and observe an increase in mean systolic and diastolic blood pressure, prolonged QTc interval, increase in pulse, QTi, and QTc, and low mean respiratory rate interval, among smokers. 
In \cite{s6}, Irfan et. al.
analyze ECG data from 5,633 patients and observe a decrease in PR interval, PR segment, and QRS duration, for smokers.  
In \cite{s7}, Ammar et. al.,
obtain 12-lead ECG from 105 subjects, with 35 being chronic e-cigarette users, and another 35 being conventional cigarette smokers. They observe an increase in mean heart rate, decrease in QRS complex duration, prolonged QT and QTc intervals, prolonged ventricular repolarization indices (T-wave-peak to T-wave-end (Tp-e) interval, Tp-e/QT ratio, and Tp-e/QTc ratio, among chronic e-cigarette users and conventional smokers. 
Gupta et. al. \cite{s8}
performs 24-hours Holter monitoring on 60 male smokers and tobacco chewers (30 in each group) experiencing atypical chest pain, and observe an increase in the heart rate in both smokers and tobacco chewers, and an increase in supraventricular ectopics and a decrease in heart rate variability for smokers.
Finally, \cite{s9} obtains ECG recordings from 37 chronic tobacco cigarette smokers and 43 chronic e-cigarette users, and compute Tp-e, Tp-e/QT ratio, and Tp-e/QTc ratio, in order to quantify the adverse effects of Nicotine on ventricular repolarization.


\section{The Methodology and the Measurement Equipment}
\label{sec:sys}
\par

\subsection{The Proposed Framework}
\begin{figure*}[h]
\centering
\includegraphics[width=0.95\textwidth]
{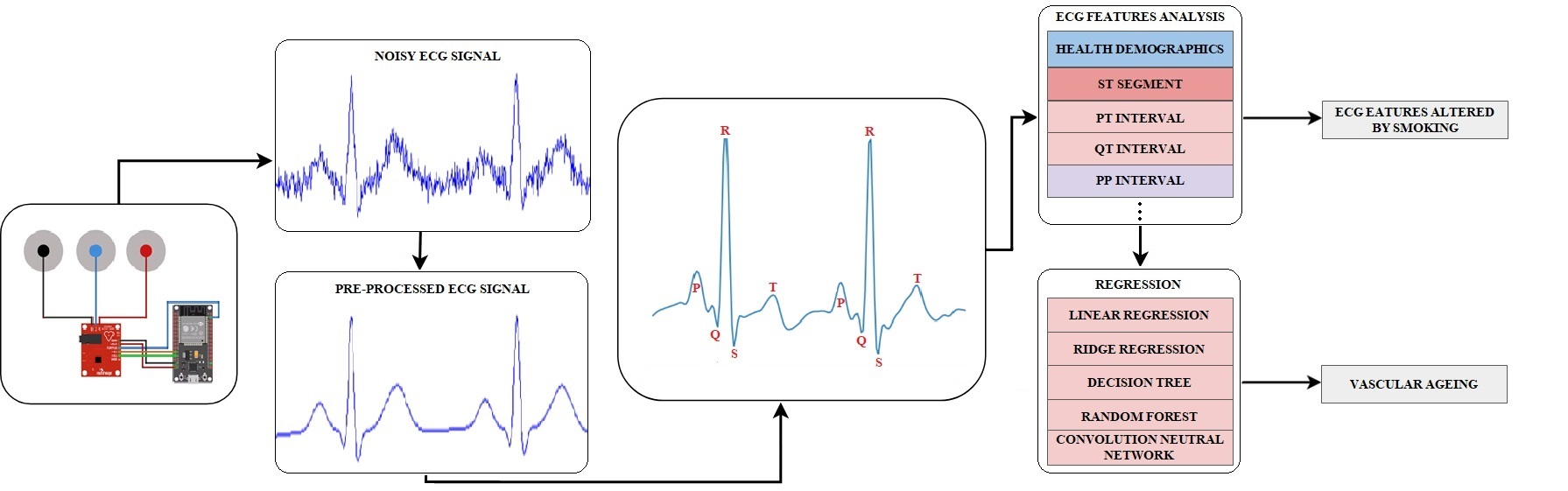}
\caption{The proposed method for vascular ageing prediction and for studying smoking-induced changes in single-lead ECG.}
\centering
\label{fig:ourmethod}
\end{figure*}

Fig. \ref{fig:ourmethod} provides a detailed overview of our proposed method. As can be seen, we utilize our custom low-cost single-lead ECG sensor module in order to construct a labelled dataset. We then do pre-processing of the noisy ECG data, followed by segmentation, followed by feature extraction. Eventually, we pass a total of 21 features (13 ECG-based features and 8 other features) to a number of ML algorithms, a 1D CNN model, and the ResNet18 model which do vascular ageing prediction. We also do transfer learning whereby we pre-train our models on a public PPG dataset, and later, fine-tune and evaluate them on our unsegmented ECG dataset. In addition, we utilize the extracted features to identify smoking-induced changes in ECG. 

\par

\subsection{The Low-cost Single-Lead ECG Module}

We utilize the AD8232 chip which is a low-cost single-lead ECG sensor (with a price tag of 5 USD). This ECG sensor requires a DC voltage in the range of 2-3.5 volts, and comes with three electrodes which enable it to capture the tiny voltages generated by the human heart, at a rate of 100 Hz. For systematic data acquisition, the AD8232 ECG sensor is interfaced with the ESP32-WROOM micro-controller board (developer version). Further, a 3.7 volts lithium-ion battery is used to power the ESP32 board, the ECG sensor, and TP4056 module (basically, a battery management system). This way, we realize a very cost-effective single-lead ECG sensor module (with a price tag of 10 USD). We connect and solder all the aforementioned hardware components on a veroboard, and then place it in an acrylic body designed via the Solidworks tool. We program the ESP32-WROOM and AD8232 ECG sensor using the Arduino IDE with the objective to read and log the ECG waveform for the purpose of our custom dataset construction. 


	

\section{Data Collection, Data Pre-Processing and Feature Extraction}

\label{sec: data collection}

\subsection{Data Collection}

\begin{figure}[h]
    \centering
    \includegraphics[width=3.45in]
    {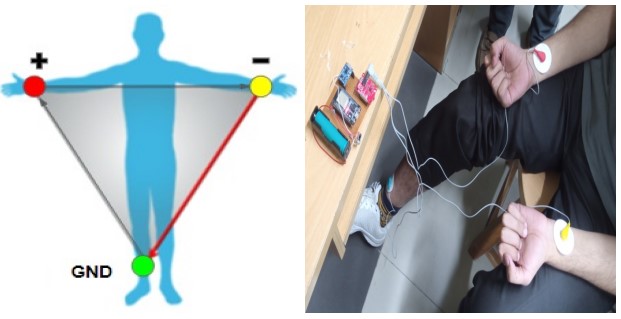}
    \caption{The three electrodes are placed such that together they form the einthoven triangle to acquire the lead I of the ECG (left subplot). The right subplot shows a snippet from our data collection campaign.}
    \label{fig:einthoven-data-cllection}
\end{figure}

Fig. \ref{fig:einthoven-data-cllection} (left subplot) shows the theoretical configuration of a three-electrode single-lead ECG sensor module, that forms the einthoven triangle for the purpose of acquiring the lead I of the ECG signal. Fig. \ref{fig:einthoven-data-cllection} (right subplot) shows a screenshot from our data collection campaign where it can be seen that the positive (red) electrode was placed on the right wrist, the negative (yellow) electrode was placed on the left wrist. Finally, the ground (green) electrode was placed on the right leg (though it is not visible in Fig. \ref{fig:einthoven-data-cllection}). 

We utilized our custom module to collect single-lead ECG data from 42 apparently healthy subjects (26 male, 16 female) of age 18 years to 30 years, on campus\footnote{This work was approved by the ethical institutional review board (EIRB) of Information Technology University (ITU), Lahore, Pakistan.}. Each subject contributed to 3 minutes worth of data, leading to a dataset having 126 minutes (7560 seconds) of single-lead ECG data. Among the male subjects, 16 were smokers, while 10 were non-smokers. Similarly, 4 female subjects were smokers, while 12 were non-smokers. All the smoker subjects in this study are light but habitual smokers, i.e., they smoke only 1-2 cigarettes per day, but have been doing it on a regular basis since last 2-3 years. For the ground truth for the vascular ageing prediction problem, we utilized chronological age as a proxy for the vascular age (which is a reasonable assumption keeping in mind that we collected data from apparently healthy young subjects only) \cite{aziz2021heartbeat},\cite{liu2018support}. In addition to acquiring the ECG traces of the subjects, we also recorded the relevant metadata such as age, height, weight, heart disease history in the family, smoking habits, sleep duration, blood pressure (systolic and diastolic), body mass index (BMI), and the resting heart rate.

\subsection{Data Pre-Processing}
\label{sec: preprocessing}
We first implemented a Butterworth low-pass filter with a cutoff frequency of 18 Hz and order 3, to denoise the ECG traces. We then removed the baseline wander (via a median filter), followed by the z-score normalization of the clean ECG traces. The onset of the P wave was used to detect the start of an ECG cycle (i.e., the PQRST wave). Anomalies in the ECG traces (that arise due to motion artifacts and due to improper placement of electrodes and cause a streak of zeros, not a number (NaN) values, or abnormal patterns) were manually identified and removed, from the onset of the P wave in that PQRST interval to the onset of the P wave in the next interval.

\subsection{Creation of 2nd Dataset via Segmentation}
Inline with the established practice of data augmentation through segmentation in the field of supervised learning, each of the 42 examples/recordings were further divided into a number of five-second long (overlapping) segments with a stride of one second. This led to another labeled dataset with 6,131 data points. Table \ref{tab:1} outlines the distribution of the two custom datasets, i.e., the original/unsegmented dataset, and segmented dataset.

\begin{table}[h]
\begin{center}

\begin{tabular}{c*{5}{c}}
\\\hline

 & Unsegmented & Segmented \\\hline\hline

Male & 26 & $3455$    \\\hline
Female & 16& $2676$  \\\hline
Total & 42 & $6131$  \\\hline
Male Smokers & 16 & $2111$   \\\hline
Female Smokers & 4 & $734$   \\\hline
Total Smokers & 20 & $2845$  \\\hline
Male Non-Smokers & 10 & $1344$   \\\hline
Female Non-Smokers & 12 & $1942$   \\\hline
Total Non-Smokers & 22 & $3286$  \\\hline

\end{tabular}
\caption{ Key statistics of the two custom datasets. }
\label{tab:1}
\end{center}
\end{table}

\subsection{Feature Extraction}
\label{sec: feature extraction}
\begin{figure}[!htbp]
\centering
\includegraphics[width=3.6in]
{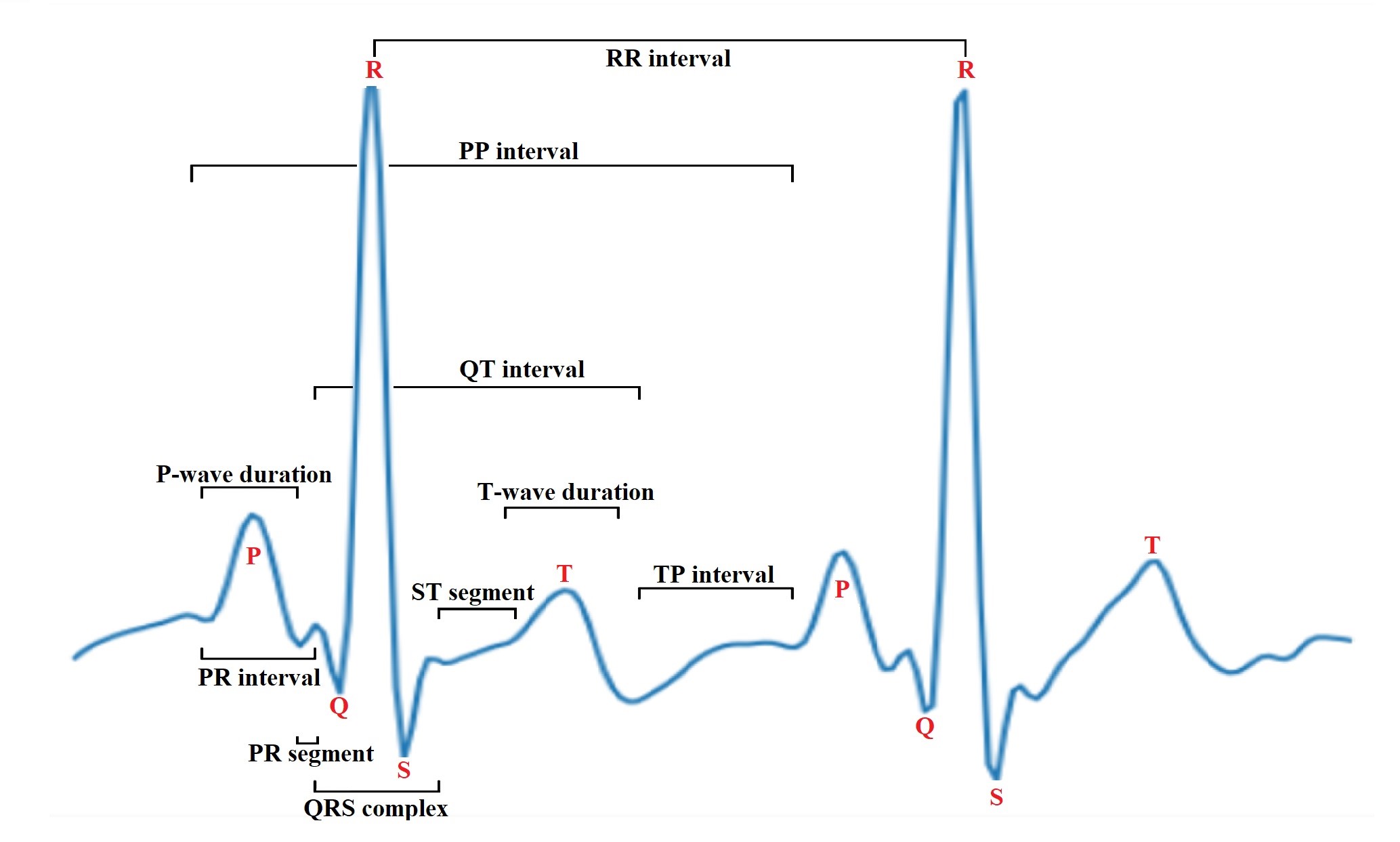}
\caption{One cardiac cycle of a typical ECG signal annotated with many intervals (features) that have clinical significance.}
\centering
\label{Figure: comp}
\end{figure}

It is well-known that the ECG signal (with P,Q,R,S,T points marked) contains a wealth of information that has clinical significance to diagnose a number of heart abnormalities, including the condition of the heart, i.e., vascular ageing. Thus, the feature extraction from ECG involves the detection of P wave, QRS complex, T wave, as well as the measurement of the duration of a number of intervals between the marked points, e.g., RR interval (see Fig. \ref{Figure: comp} that shows one typical cardiac cycle with various intervals annotated). 


We extracted a number of features, e.g., RR interval, QT interval, P wave duration, PP interval, TP interval, PR interval, T wave duration, ST segment, QRS complex duration, PR segment, etc., from both segmented ECG dataset and unsegmented ECG dataset (using the Neurokit2 library). The extraction process involved applying a discrete wavelet transform to the ECG signals, which helped focus on the various peaks in the PQRST wave (i.e., one cardiac cycle) and the corresponding intervals. 

Computing the various intervals allowed us to compute a few derived features as well. For example, we utilized the QT interval and RR interval to compute the corrected QT interval (QTc) as follows: $QTc = \frac{QT}{\sqrt{RR}}$. Further, using the RR interval, we derived the root mean square of successive differences (RMSSD) (which measures the variability between adjacent RR intervals), and the standard deviation of RR intervals (SDNN) (which quantifies the dispersion of the RR intervals).

We make the following observation regarding the segmented dataset. Recall that each ECG segment is of duration 5 seconds, thus, each segment typically consists of about 5-10 cardiac cycles. Therefore, though we initially did feature extraction on a cardiac-cycle basis, we then averaged all the 13 extracted features across a segment in order to achieve reliable and stable features. 

Table \ref{tab:table2.3} presents the values of the extracted features for one example from each of the two custom datasets (i.e., segmented and unsegmented), for illustration purposes. We observe that the values of the extracted features lie within the nominal range specified by the American heart association for the healthy young people \cite{ref35}.

\begin{table}[h]
\begin{center}
\begin{tabular}{c*{2}{c}}
\hline
\textbf{ECG Features} & \textbf{Segmented} & \textbf{Unsegmented} \\\hline\hline
RR Interval & $712.32$ & 701.15 \\\hline
RMSSD & $65.59$ & 26.78 \\\hline
SDNN & $52.40$ & 26.12 \\\hline
QT Interval & $362.33$ & 281.24 \\\hline
P-wave Duration & $101.47$ & 75.7 \\\hline
PP Interval & $777.88$ & 657.02 \\\hline 
PT Interval & $211.74$ & 421.69 \\\hline
PR Interval & $135.79$ & 230.45 \\\hline
T-wave Duration & $176.13$ & 129.21 \\\hline
ST Segment & $59.36$ & 204.78 \\\hline
QRS Complex Duration & $126.84$ & 76.48 \\\hline
Corrected QT Interval & $429.31$ & 410.14 \\\hline
PR Segment & $34.32$ & 30.29 \\\hline
\end{tabular}
\end{center}
\caption{ An illustration of the feature extraction process (i.e., one example from each of the two datasets, i.e., segmented and unsegmented datasets). Values are in milli sec. }
\label{tab:table2.3}
\end{table}

\section{Machine, Deep and Transfer Learning Models Implemented}

Keeping in mind the very small size of our custom ECG dataset (consisting of 42 subjects only), the high-complexity AI models such as transformers, deep neural networks, large language models are ruled out. This leaves us with classical ML models which typically require a few hundred of examples in order to learn the regression problem at hand.  

\subsection{Machine and deep learning models} 

{\it Machine learning models:} 
For the vascular ageing prediction problem, we implemented the following ML models: linear regression, ridge regression (linear regression with regularization), decision tree regression, and random forest regression (an ensemble of decision trees). 

{\it Deep learning models:} 
For the vascular ageing prediction problem, we implemented a 1D CNN (using the Keras library). The model utilizes 32 filters and a kernel size of 3, followed by a ReLU activation function. In the fully connected layers, a dense layer with 128 neurons and a ReLU activation are employed, followed by a dropout layer with a rate of 0.2 to mitigate overfitting. Lastly, in the output layer, a dense layer with a single neuron and a linear activation function is used for regression. The model is compiled using the Adam optimizer, with 220 epochs and a batch size of 23. 


\subsection{Transfer learning}
For the vascular ageing prediction problem, we implemented two transfer learning models, a ResNet18 model for segmented ECG dataset, and a random forest model for unsegmented ECG dataset.

First, we passed the segmented ECG dataset through ResNet18, a popular CNN model in computer vision community, in order to do transfer learning. 
The model has 18 layers, and takes as input the RGB images of size 224x224 pixels. 
The structure of ResNet-18 is as follows: i) convolutional layer (7x7, stride 2), for feature extraction; ii) max-pooling (3x3, stride 2), to reduce the spatial dimensions of the feature maps; iii) residual blocks (5 groups with 2 or 3 basic blocks each) with skip connections\footnote{Each BasicBlock layer comprises a convolutional layer, followed by batch normalization, followed by a ReLU activation function.}; iv) global average pooling, to further reduce the spatial dimensions of the feature maps; v) a fully connected layer with 512 input features, followed by a ReLU activation, an output layer with a single neuron to predict the vascular age of the input sample. 

Secondly, we trained our custom-built AI models (i.e., linear regression, ridge regression, decision tree, random forest and 1D-CNN) on a photoplethysmography (PPG)-based vascular age estimation dataset of \cite{saranVApaper}. Later, we fine-tuned and tested this random forest model on our unsegmented ECG dataset that was originally pre-trained on PPG data. This approach has many different names in literature, e.g., domain adaptation, transfer learning, representation learning, meta learning, etc.

\subsection{Training of ML, DL and TL models}
The segmented ECG dataset (consisting of 6131 samples) was divided into three parts. 60\% (i.e., 3679 samples) of the dataset was used for training, 20\% (i.e., 1226 samples) for validation, and remaining 20\% (i.e., 1226 samples) for testing. For each method, the best combination of hyperparameters was identified by conducting a grid search. For transfer learning, we translated the 1D ECG segments into spectrograms and fed them to the ResNet18 model. On the other hand, for the unsegmented ECG dataset, we do $k$-fold cross-validation with $k=5$.

\section{Results}

\subsection{Performance Metrics}
For the regression problem (of vascular aging prediction), we utilize the mean squared error (MSE) and $R^2$ score as performance metrics. Note that a higher $R^2$ score means a better fit and a lower MSE means a better model predictive accuracy. 



\subsection{Vascular Ageing Prediction}

\begin{figure}
    \centering
        \begin{subfigure}{0.96\linewidth} 
            \includegraphics[width=\linewidth]{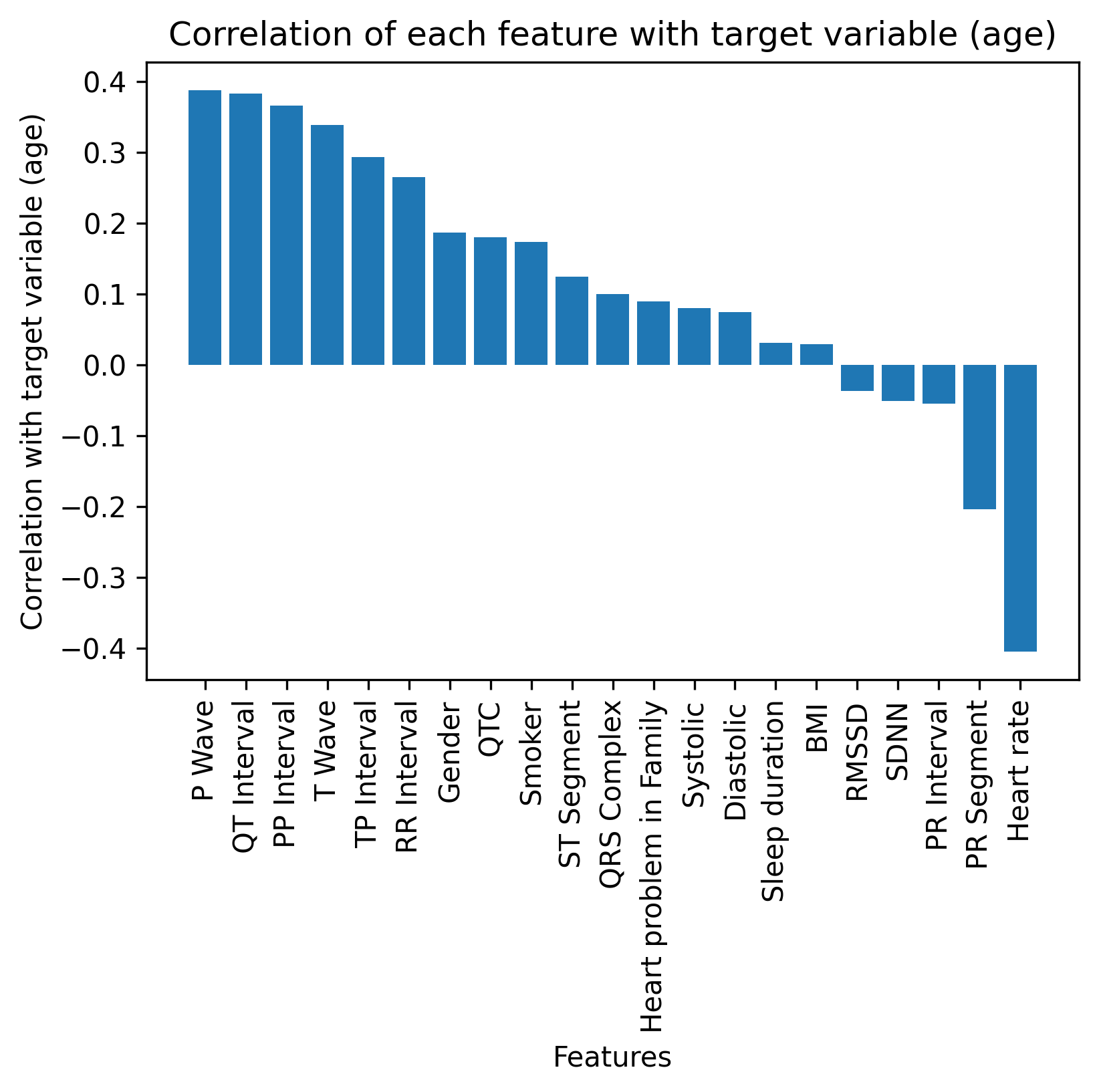}
        \caption{ Using segmented ECG dataset. }
        
    \end{subfigure}
       \hfill
    \begin{subfigure}{0.96\linewidth} 
        \includegraphics[width=\linewidth]{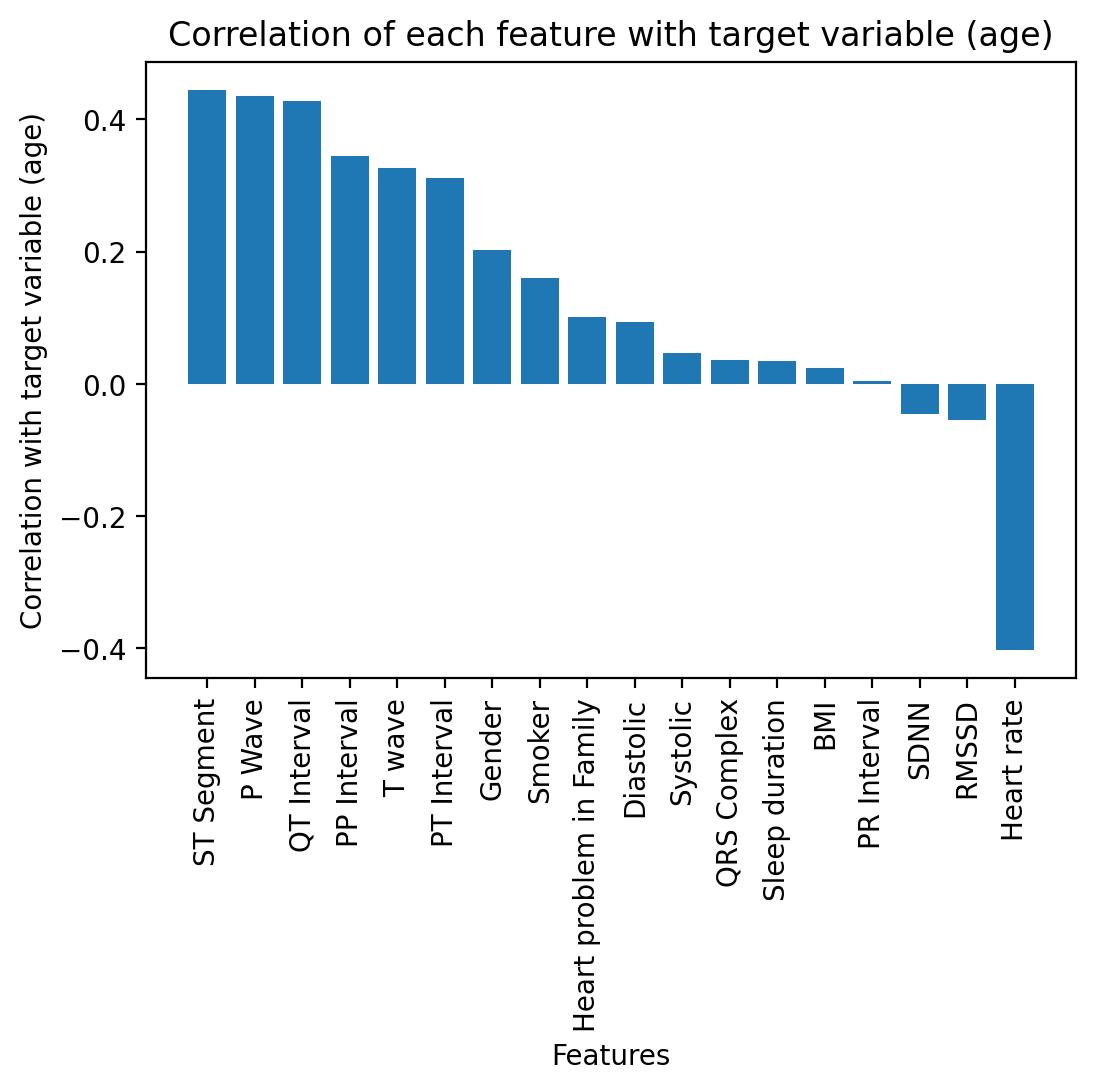}
        \caption{Using unsegmented ECG dataset.}
        
    \end{subfigure}
    
    \caption{Explainable AI based feature correlation analysis. (target variable is age).}
    \label{fig:xAI-age}
\end{figure}

{\it Explainable AI analysis:}
We begin by doing the feature correlation analysis in spirit of explainable AI (xAI), for one example from each of the two datasets, i.e., segmented ECG dataset and unsegmented ECG dataset. We learn from Fig. \ref{fig:xAI-age} that feature extraction from both segmented and unsegmented ECG datasets yields very similar results. That is, both Fig. \ref{fig:xAI-age} (a) and Fig. \ref{fig:xAI-age} (b) illustrate that the following ECG features are most relevant to the vascular ageing: heart rate, P-wave duration, QT interval, PP interval, and T-wave duration. 
Similarly, both Fig. \ref{fig:xAI-age} (a) and Fig. \ref{fig:xAI-age} (b) reveal that vascular ageing is agnostic to the following ECG and demographic features: BMI, sleep duration, RMSSD, and SDNN. 

However, we observe that despite lots of agreement between Fig. \ref{fig:xAI-age} (a) and Fig. \ref{fig:xAI-age} (b), there are some differences as well. For example, ST segment is considered as an important ECG feature by the feature extraction process done on unsegmented dataset, but not by segmented dataset. As another example, PR segment is considered as an important ECG feature by the feature extraction process done on segmented dataset, but not by unsegmented. We believe this minor discrepancy is an artifact due to small size of our custom dataset. But the overall agreement between Fig. \ref{fig:xAI-age} (a) and Fig. \ref{fig:xAI-age} (b) points to the legitimacy of the segmentation process.

\begin{table*}[h!]
\centering
\begin{tabular}{|c|c|c|c|}
\hline
AI model & Segmented (S) & Unsegmented (US) & Unsegmented+transfer learning (US+TL) \\
\hline
Linear regression & MSE=5.79; $R^2$=0.337 & MSE = 14.13; $R^2$ = -1.92 & MSE= 4.36; $R^2$= 0.45 \\
\hline
Ridge regression & MSE=5.24; $R^2$=0.37 & MSE = 18.19; $R^2$ = -2.76 & MSE = 4.39; $R^2$ = 0.45 \\
\hline
Decision tree & MSE=0.258; $R^2$=0.97 & MSE = 6.13; $R^2$ = -0.26 & MSE= 4.28; $R^2$=0.459 \\
\hline
\rowcolor{pastelblue}\bf Random forest & MSE=0.07; $R^2$=0.99 & MSE = 3.56; $R^2$ = 0.26 & MSE = 0.99; $R^2$= 0.87 \\
\hline
1D-CNN & MSE=0.251; $R^2$=0.971 & MSE = 4.24; $R^2$ = 0.12 & MSE= 28.2; $R^2$= -2.564 \\
\hline
\end{tabular}
\caption{Performance of our AI models for vascular ageing prediction, across the three learning scenarios.}
\label{table:ourAImodels}
\end{table*}

Table \ref{table:ourAImodels} provides a detailed performance comparison of all the AI models across the three distinct learning scenarios that we have implemented for vascular age prediction. Specifically, we have evaluated our AI models: 1) on the segmented ECG dataset, 2) on the unsegmented (original) ECG dataset as is, and 3) for the scenario where our models are first pre-trained on the PPG dataset of \cite{saranVApaper}, and later fine-tuned and tested on our custom unsegmented ECG dataset. 
For the first learning scenario where we evaluate all the models on segmented ECG dataset, the linear ML models, i.e., linear regression model and its variant ridge regression perform poorly which hints at a possible non-linear relationship between the ECG-based features and the vascular age. This hypothesis is later verified as the decision tree model and random forest model (an ensemble of decision trees) perform very well with a smaller MSE and higher $R^2$ score. On the deep learning front, the 1D CNN also performs well but at a cost of more computational complexity. Next, for the second learning scenario where we evaluate all the models on unsegmented ECG dataset, we notice that the linear regression and ridge regression collapse because of the small size of the unsegmented dataset and because of non-linear relationship between the ECG-based features and the vascular age. Finally, for the third scenario where we fine-tune and test all the models (that were pre-trained on PPG dataset of \cite{saranVApaper}) on unsegmented ECG dataset, all models perform decently, but the 1D-CNN also collapses. This is perhaps due to the small size of our unsegmented ECG dataset that we use for fine-tuning purposes. All in all, the random forest model comes out as a winner across all three learning scenarios. That is, the random forest model achieves an MSE of 0.07 and $R^2$ of 0.99, MSE of 3.56 and $R^2$ of 0.26, MSE of 0.99 and $R^2$ of 0.87, for segmented ECG dataset, for unsegmented ECG dataset, and for transfer learning scenario, respectively. The superior performance of the random forest is because of the following: i) it is an ensemble method so it produces more reliable aggregate decisions due to so many decision tree at the backend, and ii) its complexity is well-matched with the size of our unsegmented ECG dataset. 
  

\subsection{ Comparison with related works on vascular ageing prediction}

Table \ref{tab:comparisonwithsota} provides a detailed performance comparison of the top performing random forest model (for all three learning scenarios) against the related works on vascular ageing estimation. Table \ref{tab:comparisonwithsota} highlights that our random forest model provides MSE and $R^2$ performance that is at par with the performance of 12-lead ECG-based state-of-the-art methods for vascular aging prediction.

\begin{table}[!ht]
    \centering
    \begin{tabular}{p{0.4in}|p{0.17in}|p{0.4in}|p{0.5in}|p{0.27in}|p{0.5in}}
    \hline
     \textbf{Authors} & \textbf{Year} & \textbf{Waveform} & \textbf{Model implemented} & \textbf{Age Range} & \textbf{Performance Metric}\\ \hline

     Starc et al. \cite{starc2012can} & 2012 & 12-lead ECG & Multiple Linear Regression & 3-100 & R\textsuperscript{2}=0.76 \\ \hline
     
     Lima et al. \cite{lima2021deep} & 2021 & 12-lead ECG & Residual Network & 20-80 & MAE=8.83 \\ \hline
     
     Chang et al. \cite{chang2022electrocardiogram} & 2022 & 12 Lead ECG & Deep Learning & 20-80 & MAE=6.899 R\textsuperscript{2}=0.822 \\ \hline

     Zvuloni et al. \cite{zvuloni2023merging} & 2023 & 12-lead ECG & FE and DL & 20-85 & MAE\textsubscript{FE}=10.6 R\textsuperscript{2}\textsubscript{FE}=0.60 MAE\textsubscript{DL}=6.32 R\textsuperscript{2}\textsubscript{DL}=0.83  \\ \hline
     
     \textbf{Ours} & \textbf{2024} & \textbf{1-lead ECG} & \textbf{Random Forest (S)} & \textbf{18-30} & \textbf{MSE=0.07; R\textsuperscript{2}=0.99} \\ \hline

     \textbf{Ours} & \textbf{2024} & \textbf{1-lead ECG} & \textbf{Random Forest (US)} & \textbf{18-30} & \textbf{MSE=3.56; R\textsuperscript{2}=0.26} \\ \hline

     \textbf{Ours} & \textbf{2024} & \textbf{1-lead ECG} & \textbf{Random Forest (US+TL)} & \textbf{18-30} & \textbf{MSE=0.99; R\textsuperscript{2}=0.87} \\ \hline
 
    \end{tabular}
    \caption{MSE and $R^2$ accuracy of our ML, DL and TL models and their comparison with the previous work.}
    \label{tab:comparisonwithsota}
\end{table}

\begin{figure*}
    \centering
        \begin{subfigure}{0.94\linewidth} 
        \includegraphics[width=\linewidth]{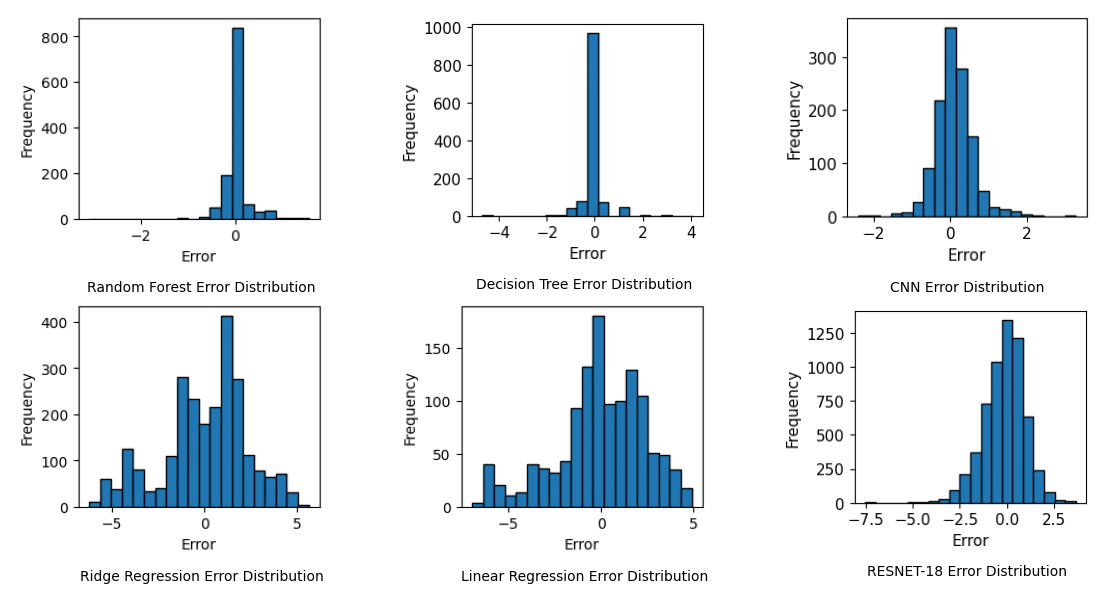}
        \caption{ Using segmented ECG dataset. }
        
    \end{subfigure}
       \hfill
    \begin{subfigure}{0.94\linewidth} 
        \includegraphics[width=\linewidth]{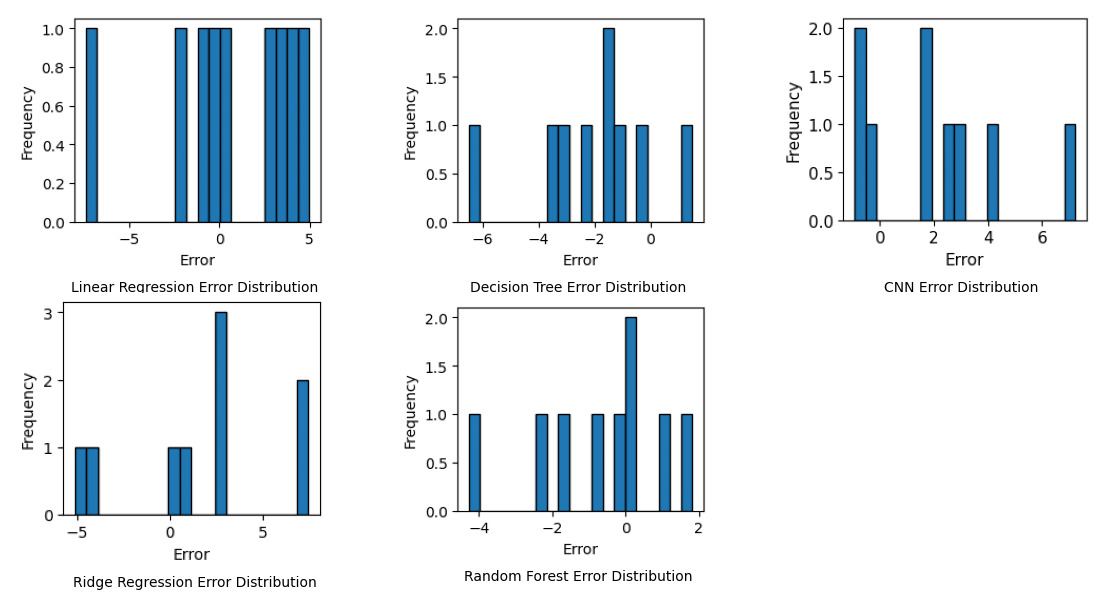}
        \caption{Using unsegmented ECG dataset.}
        
    \end{subfigure}
    
    \caption{Error Distribution of our ML, DL and TL models.}
    \label{fig:histograms}
\end{figure*}

Fig. \ref{fig:histograms} plots the histograms of the error in vascular aging prediction for all models we have implemented, for both segmented and unsegmented ECG datasets. For the segmented ECG dataset, we observe that the random forest, the decision tree, and 1D CNN have the smallest error spread among all, as expected. For the unsegmented ECG dataset, the histograms don't reveal much about the error spread and error trends, due to small size of the dataset. 

\subsection{Effect of Light-but-Habitual Smoking on ECG}

\begin{figure}
    \centering
        \begin{subfigure}{0.96\linewidth} 
        \includegraphics[width=\linewidth]{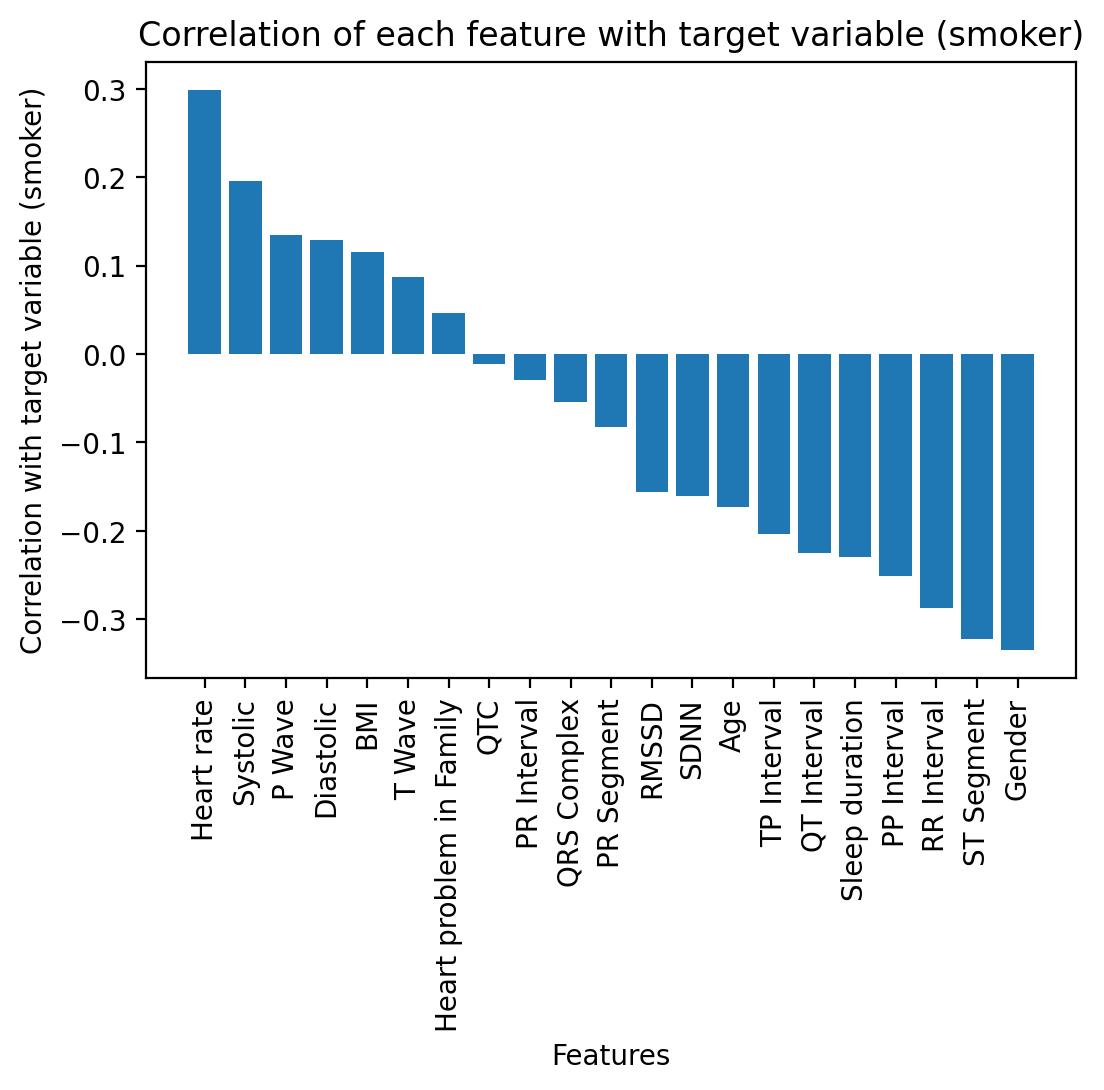}
        \caption{ Using segmented ECG dataset. }
        
    \end{subfigure}
       \hfill
    \begin{subfigure}{0.96\linewidth} 
        \includegraphics[width=\linewidth]{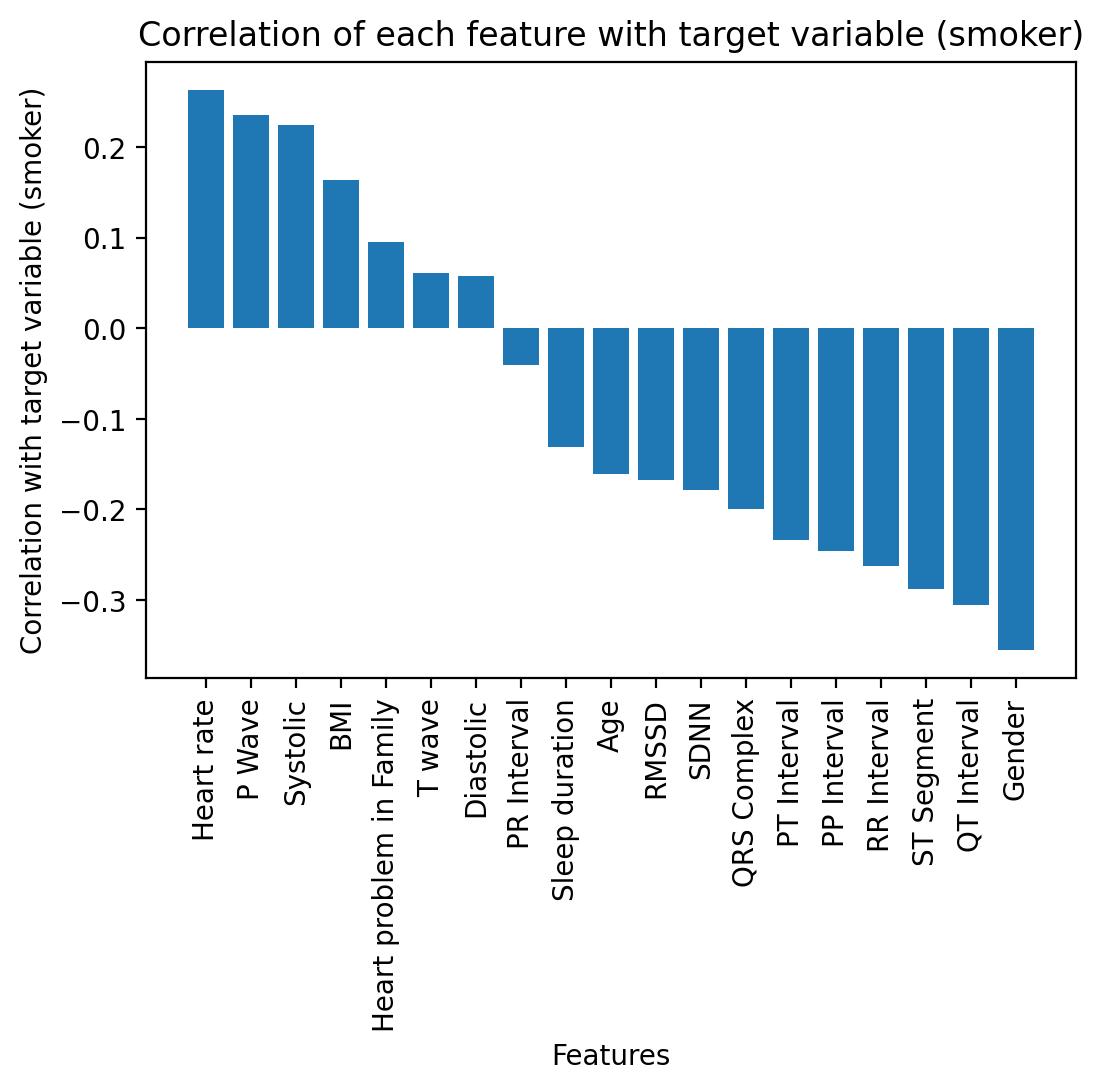}
        \caption{Using unsegmented ECG dataset.}
        
    \end{subfigure}
    
    \caption{Explainable AI based feature correlation analysis. (target variable is smoking).}
    \label{fig:xAI-smoking}
\end{figure}

{\it Explainable AI analysis:}
We again do feature correlation analysis in spirit of xAI, by taking one example from each of the two datasets, i.e., segmented and unsegmented datasets. We learn from Fig. \ref{fig:xAI-smoking} that feature extraction from both datasets, i.e., segmented ECG dataset and unsegmented ECG dataset, leads to findings that are mostly in agreement. 
That is, the following ECG and demographic features show the highest correlation with smoking: heart rate, P wave duration, systolic blood pressure, BMI, for both datasets. 
Further, the following ECG and demographic features show the highest negative correlation with smoking: gender, ST segment, RR interval, PP interval, PT (or TP) interval, for both datasets. 
However, we note that there is a slight disagreement between the two datasets regarding a few features, e.g., regarding QRS complex duration whereby the unsegmented dataset considers it important, while segmented dataset doesn't. Similarly, sleep duration is important for segmented dataset but not for unsegmented dataset. Furthermore, QT interval is deemed more important by unsegmented dataset, while it is considered less important by segmented dataset. 
We believe this discrepancy is due to small size of our custom dataset. Overall, a good agreement between Fig. \ref{fig:xAI-smoking} (a) and Fig. \ref{fig:xAI-smoking} (b) validates the segmentation process carried out in this work. 


\begin{figure*}[h]
\centering
\includegraphics[width=7.5in]
{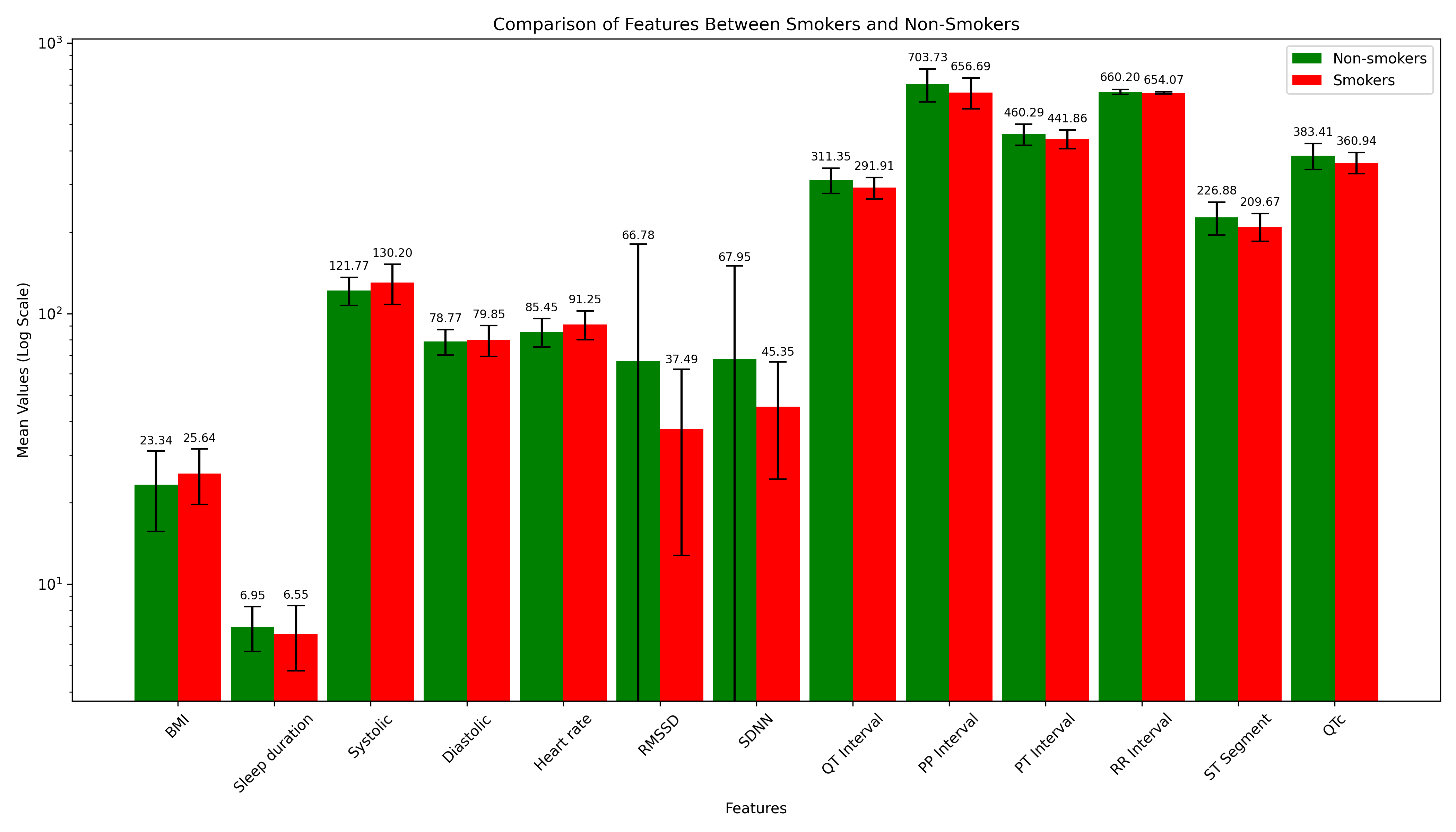}
\caption{Impact of smoking on ECG features.}
\centering
\label{fig:smoking_ecg}
\end{figure*}

Next, to analyze the effect of smoking on ECG features at an aggregate level, we did feature extraction from the full unsegmented dataset.  Fig. \ref{fig:smoking_ecg} plots the mean values and standard deviations of 11 most relevant ECG features for both smoker and non-smoker groups. We notice that the following features show noticeable changes (in terms of mean and standard deviation) across the two classes (i.e., smokers and non-smokers): RMSSD, and SDNN. This is followed by the following features: sleep duration and BMI. The following ECG features register mediocre changes across the two classes: systolic blood pressure, heart rate, QT interval, PP interval, and ST segment. On the other hand, the following ECG features remain unchanged across the two classes: diastolic blood pressure, and PT interval. Thus, even habitual-but-light smoking causes changes/abnormalities in ECG traces, as shown by the changes in the feature values in Fig. \ref{fig:smoking_ecg}. This could in term help do early prediction of the risk of cardiovascular diseases among the smokers group.

\section{Data availability statement}
The custom ECG dataset used and analysed during this research study is available from the corresponding author on reasonable request.

\section{Declarations}
\subsection{Funding and/or Conflicts of Interests/Competing Interests}
Funding: This work was supported in part by the UK Engineering and Physical Sciences Research Council (EPSRC) grants: EP/X040518/1 and EP/T021020/1.
Conflict of Interest: The authors declare that they have no conflict of interest.

\section{Conclusion}

This paper predicted the vascular age of a person and studied smoking-induced changes in ECG using a custom-built low-cost single-lead ECG sensor module. We constructed our own custom dataset by collecting ECG data and metadata from 42 apparently healthy young subjects (both smokers and non-smokers). 
We passed the segmented and unsegmented datasets to a number of AI models. We also did transfer learning whereby
we pre-trained our models on a public PPG dataset, and later, fine-tuned and evaluated them on our unsegmented ECG dataset.
The random forest model outperformed all other models and previous works by achieving a mean squared error (MSE) of 0.07 and cofficient of determination $R^2$ of 0.99, MSE of 3.56 and $R^2$ of 0.26, MSE of 0.99 and $R^2$ of 0.87, for segmented ECG dataset, for unsegmented ECG dataset, and for transfer learning scenario, respectively.
Finally, we utilized the explainable AI framework to identify the most relevant ECG features that change due to smoking.

This work is aligned with the sustainable development goals (3 and 10) of the United Nations which aim to provide low-cost but quality healthcare sensing solutions to the unprivileged (e.g., people in developing countries and people in remote areas). Furthermore, this work opens the door for innovation in wearable devices (e.g., smartwatches and wristbands), enabling accurate and real-time prediction of the vascular age using a single-lead (lead-I) ECG sensor. In addition, this work also finds its applications in the broad domain of forensic science.

Finally, we note that our dataset consists of resting-state ECG data collected from apparently healthy young subjects only. Therefore, further efforts are required to collect large amounts of data from a more diverse population (e.g., with different ages, smoking habits, cardiovascular diseases, varying time gaps between last smoking instance and the ECG measurement etc.). Such dataset will help understand the more fine-grained implications of smoking on cardiovascular health, as well as provide valuable insights about the generalization capability of the AI models in real-world settings.





\footnotesize{
\bibliographystyle{IEEEtran}
\bibliography{references}
}

\vfill\break

\end{document}